# Stereotype Threat?: Effects of Inquiring About Test Takers' Gender on Conceptual Test Performance in Physics

Alexandru Maries[1, a)] and Chandralekha Singh[1,2]

[1]*Discipline-Based Science Education Research Center (dB-SERC),* [2]*Department of Physics and Astronomy*
*University of Pittsburgh, Pennsylvania, USA*

a)hitren86@gmail.com

**Abstract.** It has been found that activation of a stereotype, for example by indicating one's gender before a test, typically alters performance in a way consistent with the stereotype, an effect called "stereotype threat." On a standardized conceptual physics assessment, we found that asking test takers to indicate their gender right before taking the test did not deteriorate performance compared to an equivalent group which did not provide gender information. Although a statistically significant gender gap was present on the standardized test whether or not students indicated their gender, no gender gap was observed on the multiple-choice final exam students took, which included both quantitative and conceptual questions on similar topics.

## INTRODUCTION AND METHODOLOGY

Prior studies have found that performance on tests can be significantly altered if a stereotype about a group is activated prior to test administration. For example, women who are told that a math test they are about to take typically shows gender differences perform significantly worse than women who are not told anything about the test [1]. This is one example of stereotype threat. Stereotype threat can create, for example, increased anxiety and reduced working memory capacity. Researchers have identified a broad array of stimuli ranging from very subtle (e.g., writing down ethnicity) to overt (e.g., explicitly reminding participants about the stereotype) that can activate a stereotype and influence performance [2]. For example, a highly cited research study [3] found that the performance of African-American students on a difficult test of verbal ability was lower when they were asked to indicate their ethnicity compared to when they were not asked for this information. This research suggests that merely asking students to indicate their ethnicity can activate negative stereotypes about ability and result in decreased performance. However, other researchers [4] have found no effects that are both statistically and practically significant resulting from asking students to indicate their ethnicity or gender on standardized tests (e.g., AP Calculus exam).

In physics, prior research has often shown a gender gap [5] (i.e., worse female performance on conceptual assessments), and sometimes even on high-stakes multiple-choice tests such as final exams [6]. In the research presented here, we investigate the gender gap both on a conceptual assessment (the Conceptual Survey of Electricity and Magnetism [7], or CSEM) and on a multiple-choice final exam for students in second semester algebra-based introductory physics courses (who were mostly biological or neuro science majors or premeds) at the University of Pittsburgh. We also investigate whether in the context of CSEM, asking students to indicate their gender activates negative stereotypes about performance for females and results in a decreased performance. Students took the CSEM both as a pre-test during the first week and as a post-test after learning the relevant concepts. On the front page of the CSEM, students were asked to indicate some demographic information and were randomly assigned to two conditions, one which included a question about gender (in checkbox format: male, female, prefer to not specify) which we refer to as "the gender salient condition" and the other which did not. It is possible that the commonly reported gender gap on standardized physics tests is partly due to the multiple choice format of the tests used, or partly due to the conceptual nature of the test questions typically used. We therefore compared the performance of males and females on the instructor-developed multiple choice final exam, both on the conceptual and on the quantitative questions (the final exam contained a roughly even mix of conceptual and quantitative questions).

# RESULTS

Figure 1a shows that the performance of males and females on the CSEM is not affected by the condition students were assigned to (gender salient, not salient) both in the pre-test and in the post-test. Furthermore, on both the pre-test and the post-test, males outperformed females ($p = 0.019$ and $p = 0.028$, respectively). On the other hand, Figure 1b shows that on the final exam, the performance of females and males is nearly identical, both on the conceptual and quantitative questions.

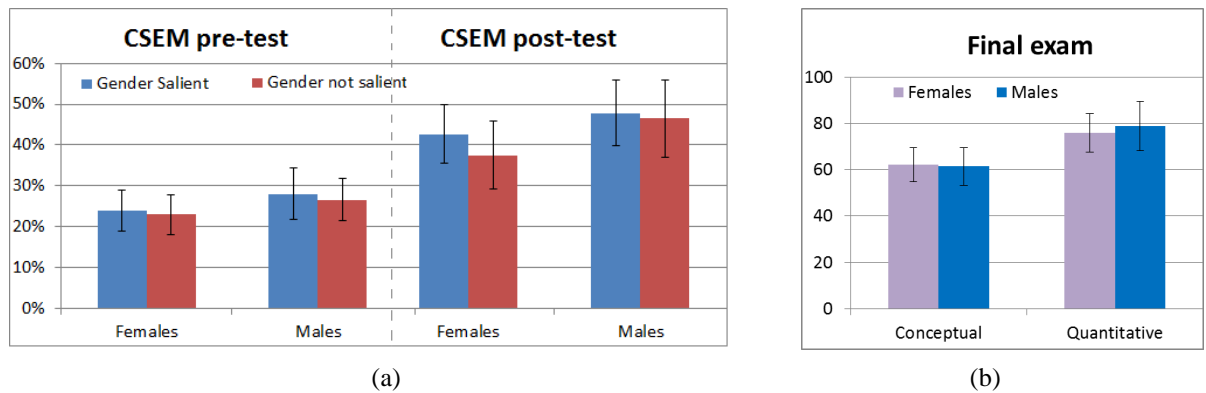

(a)                                                (b)

**FIGURE 1.** Performance on (a) the CSEM and (b) the multiple-choice final exam questions

# SUMMARY AND CONCLUDING REMARKS

In this study, asking students to indicate their gender did not alter their performance on the CSEM, consistent with prior findings [4] in a nonphysics context. It is possible that negative stereotypes about female performance in physics are activated whether or not students identify their gender before taking a test so females perform worse than males regardless of whether they are asked to write their gender. If these results are also true for high stakes standardized tests (e.g., SAT, GRE), this result may be interpreted to imply that the common practice of asking for demographic information before the test is not detrimental to groups which are stereotypically perceived as underachieving on these tests. In addition, we found the commonly reported gender gap on the CSEM, but did not find any gender gap on the multiple-choice final exam, which included both conceptual and quantitative questions on similar topics as the CSEM. The presence of a gender gap on the CSEM but not on the final exam can imply that something special about standardized tests partly accounts for the gender gap. Some have argued that standardized tests are gender biased, partly due to the masculine contexts of questions typically used in these tests, but the multiple choice questions on the final exam used in this study had similar masculine contexts to the questions on the CSEM and did not show a gender gap. We hypothesize that possible reasons for the differences between the performance on CSEM and final exam may include the fact that instructors had conducted a review session or given a practice exam before the final exam which may have alleviated some of the female anxiety about test-taking [8].

# REFERENCES

1. S. J. Spence, C. M. Steele, and D. M. Quinn, "Stereotype threat and women's math performance," J. Exp. Soc. Psychol. **35**, 4-28 (1999).
2. S. C. Wheeler and R. E. Petty, "The effects of stereotype activation on behavior: A review of possible mechanisms," Psychol. Bull. **127**(6), 797-826 (2001).
3. C. M. Steele and J. Aronson, "Stereotype threat and the intellectual performance of African Americans," J. Pers. Soc. Psychol. **69**(5), 797-811 (1995).
4. L. J. Strickler and W. C. Ward, "Stereotype threat, inquiring about test takers' ethnicity and gender, and standardized test performance," J. Appl. Soc. Psychol. **34**(4), 665-693 (2004).
5. M. Lorenzo, C. Crouch, and E. Mazur, "Reducing the gender gap in the physics classroom," Am. J. Phys. **74**(2), 118-122 (2006).
6. A. Miyake, L. E. Kost-Smith, N. D. Finkelstein, S. J. Pollock, G. L. Cohen, and T. A. Ito, "Reducing the gender achievement gap in college science: A classroom Study of Values Affirmation," Science **330**, 1234-1237 (2010).
7. D. P. Maloney, T. L. O'Kuma, C. Hieggelke, and A. Van Heuvelen, "Surveying students' conceptual knowledge of electricity and magnetism," Am. J. Phys. **69**(7), S12-S23 (2001).